\documentclass{article}
\usepackage{spconf, amsmath, graphicx}
\usepackage{float}
\usepackage{subfigure}
\usepackage{cite}
\usepackage[urlcolor=blue]{hyperref}

\makeatletter
\newcommand*\bigcdot{\mathpalette\bigcdot@{.5}}
\newcommand*\bigcdot@[2]{\mathbin{\vcenter{\hbox{\scalebox{#2}{$\m@th#1\bullet$}}}}}
\makeatother

\title{Context-aware Coherent Speaking Style Prediction with Hierarchical Transformers for Audiobook Speech Synthesis}
\name{Shun Lei$^{1\ddagger}$\thanks{$^{\ddagger}$Work conducted when the first author was intern at Xverse Inc.}, Yixuan Zhou$^{1\dagger}$, Liyang Chen$^1$, Zhiyong Wu$^{1,2,4*}$\thanks{$^{\dagger}$ Equal contribution. $^{*}$ Corresponding author.},  Shiyin Kang$^{3*}$, Helen Meng$^4$}
\address{
    $^1$ Shenzhen International Graduate School, Tsinghua University, Shenzhen $^2$ Peng Cheng Lab, Shenzhen\\
    $^3$ XVerse Inc., Shenzhen $^4$ The Chinese University of Hong Kong, Hong Kong SAR\\
    \small{
        \{leis21, zhouyx20, cly21\}$@$mails.tsinghua.edu.cn, 
        zywu$@$sz.tsinghua.edu.cn,
        kangshiyin$@$xverse.cn,
        hmmeng$@$se.cuhk.edu.hk
    }
}

\begin{document}
\ninept
\maketitle
\begin{abstract}
Recent advances in text-to-speech have significantly improved the expressiveness of synthesized speech.
However, it is still challenging to generate speech with contextually appropriate and coherent speaking style for multi-sentence text in audiobooks.
In this paper, we propose a context-aware coherent speaking style prediction method for audiobook speech synthesis.
To predict the style embedding of the current utterance, a hierarchical transformer-based context-aware style predictor with a mixture attention mask is designed, considering both 
text-side context information and speech-side style information of previous speeches.
Based on this, we can generate long-form speech with coherent style and prosody sentence by sentence.
Objective and subjective evaluations on a Mandarin audiobook dataset demonstrate that our proposed model can generate speech with more expressive and coherent speaking style than baselines, for both single-sentence and multi-sentence test\footnote{Speech sample: \href{https://thuhcsi.github.io/icassp2023-coherent-tts}{https://thuhcsi.github.io/icassp2023-coherent-tts}}.
\end{abstract}

\section{Introduction}
Text-to-speech (TTS) aims to generate intelligible and natural speech from text.
With the development of deep learning, 
now TTS models 
can produce high-quality and natural speech with a neutral speaking style \cite{tacotron2, fastspeech2}.
But the speaking style with limited expressiveness still remains a clear gap between the synthesized speech and human recordings, thus can no longer meet the increasingly high demands of customers in some scenarios like audiobooks.
It is still challenging to achieve audiobook speech synthesis, since the corpus usually has rich and complex stylistic variations affected by multiple factors (including contextual information, 
speaker's intention, 
etc.), which increases the difficulty of style modeling.

To model such expressive speaking style, the text-predicted global style token (TP-GST) \cite{tpgst} firstly introduces the idea of
predicting style embedding 
from input text,
which can generate voices with more pitch and energy variation than baseline Tacotron \cite{tacotron}.
Considering that style and semantic information of utterance are closely related, the text embeddings derived from pre-trained language models, e.g., Bidirectional Encoder Representations from Transformer (BERT) \cite{bert}, have been incorporated to TTS models for better predicting the style representation of speech \cite{wsv, ren2022prosospeech, wsvs}.
Furthermore, the diverse style variations of the utterance can also be affected by its neighboring sentences, especially for successive multi-sentence text in audiobooks \cite{survey,cole2015prosody}.
In this regard, some studies employ context-aware text embeddings obtained by feeding BERT with cross-sentence to improve the prosody generation for each utterance  
\cite{xu2021improving, nakata11audiobook}. 
To utilize the inherent hierarchical structure in a wide range of context,
\cite{lei2022towards} designs a hierarchical context encoder (HCE) to gather the contextual information from a few past and future adjacent sentences besides the current one, which shows great improvement in expressive speaking style prediction.

However, when combining the expressive utterances synthesized by HCE into a long-form audiobook speech, it still suffers from a certain degree of unnaturalness.
For example, the prosody varying greatly from one sentence to the next, resulting in unsmooth transitions.
This can be explained by the fact that HCE predicts the speaking style of each utterance in isolation, ignoring the prosodic coherence among them.
In addition, several researches 
have proved the existence of super-sentential prosody patterns in discourse segments, e.g., pitch reset and declination, and in different speech styles \cite{farrus2016paragraph ,peiro2018paragraph}.
Thus, it is important to generate speech with both contextually appropriate and coherent speaking style for multi-sentence in audiobook.
An intuitive way is to implement a long-form TTS model, which takes phoneme sequence of multiple sentences or even a paragraph as input, and generate  corresponding mel-spectrograms directly \cite{makarov22_interspeech, xue2022paratts}.
Nevertheless,
it requires large memory growth corresponding to the input length and
still suffers performance degradation when synthesizing extra-long text unseen in the training corpus.

In this paper, we propose a 
context-aware coherent speaking style prediction method
for audiobook speech synthesis.
To predict the style embedding of the current sentence,
a context-aware style predictor is designed considering
both text-side context information and speech-side style information of previous speeches.
Specifically, it is a hierarchical transformer architecture with a 
mixture attention mask, which can  better learn the relationship between context and style information.
Meanwhile, a style extractor is used to derive the style embedding from the mel-spectrogram of each utterance, which explicitly guides the training of the context-aware style predictor.
In this way, we can generate long-form speech with coherent style and prosody sentence by sentence.
Objective and subjective evaluations on a 
Mandarin audiobook dataset demonstrate that the proposed model outperforms all baseline approaches in terms of expressiveness and coherence of synthesized speech, for both single-sentence and multi-sentence test.

\section{methodology}
\label{sec:method}
\label{sec:intro}

\begin{figure}[!tb]
	\centering
	\includegraphics[width=0.70\linewidth]{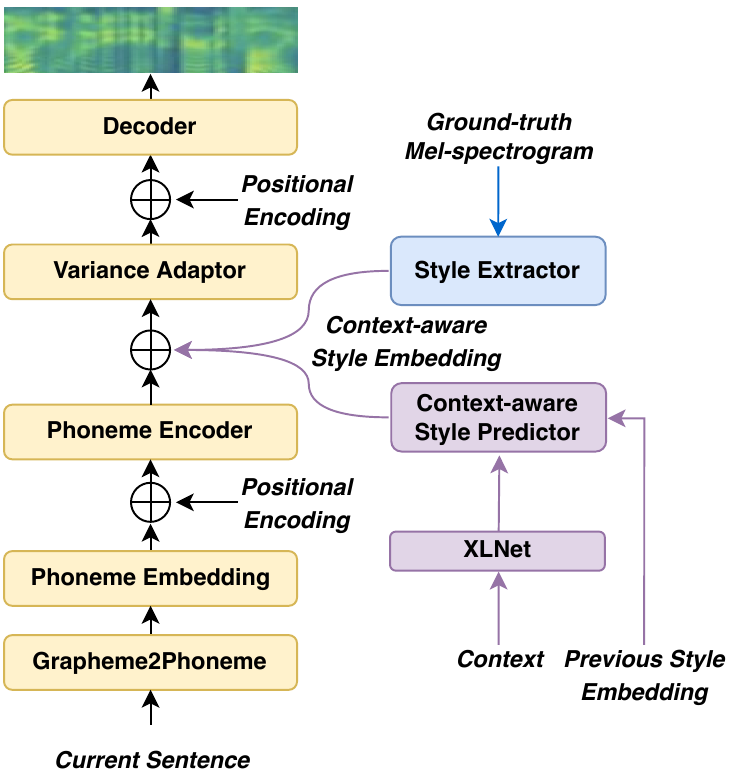}
	\caption{The overall architecture of the proposed model}
	\label{fig:model}
\end{figure}

The architecture of our proposed model is illustrated in Fig.\ref{fig:model}.
It consists of three major parts: a style extractor, a context-aware style predictor, and a sequence-to-sequence acoustic model based on FastSpeech 2 \cite{fastspeech2}.
The style extractor is used to derive the style embedding from speech.
The context-aware style predictor is used to predict the 
style embedding for each sentence,
with its context and style embeddings from its previous speeches as inputs.
Then the style embedding provided by the extractor or the predictor is replicated to the phoneme level added to the output of the phoneme encoder in the acoustic model to synthesize expressive and coherent speech.
The details of each component are as follows.

\subsection{Style Extractor}
Inspired by the success of the Global Style Token (GST) \cite{gst} model in style transfer, we introduce a style extractor to derive the 
utterance-level
style embedding from speech in an unsupervised manner.
This style extractor comprises a reference encoder and a style token layer.
Each of these modules has the same architecture and hyperparameters as the original GST.
The mel-spectrogram is passed to the style extractor whose output is regarded as the style embedding of this utterance.

\subsection{Context-aware Style Predictor}
To better utilize the text-side context information and the speech-side style information, we propose a context-aware style predictor based on HCE. 
The context information is denoted as a fixed number of the past and future adjacent sentences besides the current one.
The style information is denoted as the style embeddings extracted from the speeches of the past sentences, because only previous speeches can be available to the model during inference.

\begin{figure}[!tb]
	\centering
	\includegraphics[width=0.70\linewidth,height=0.7\linewidth]{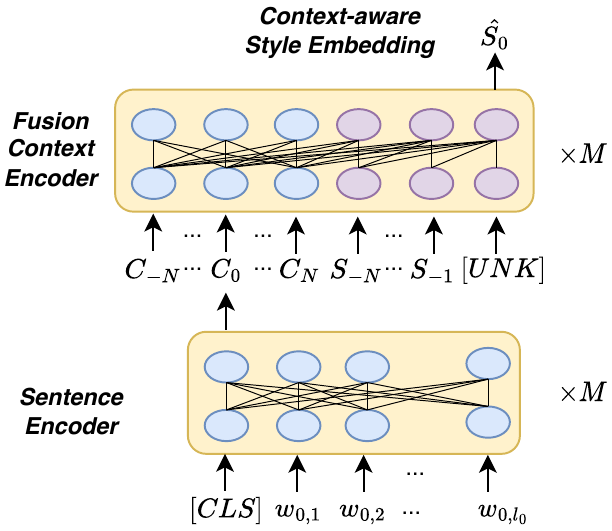}
	\caption{The Context-aware Style Predictor}
	\label{fig:predictor}
\end{figure}

Let $N$ be the number of sentences considered in the past or future.
We firstly concatenate all the $2N+1$ sentences to form a long text sequence as the context. 
Since that XLNet \cite{xlnet} can directly process longer text even the paragraphs without length limitation, 
we employ XLNet to derive word-level text embeddings from this sequence, denoted as $(w_{-N,1},w_{-N,2},\dots,w_{N,l_N})$, where $l_i$ is the number of words in sentence $i$.
To capture the inherent structural information among sentences, we introduce a hierarchical framework composed of a sentence encoder and a fusion context encoder, as shown in Fig.\ref{fig:predictor}.
The two encoders have similar architectures composed of a stack of Transformer blocks.
Firstly, the word embedding sequence of each sentence is passed to the sentence encoder, with a special [CLS] token added in front. 
Then the sentence encoder transforms the input into a hidden state sequence by exploring the low-level word meanings within a sentence.
The first hidden state corresponding to the [CLS] token is regarded as the context token of the sentence $i$, denoted as $C_i$.
After that, the context token sequence $C_{-N},C_{1-N},\cdots,C_{N}$  is passed to the fusion context encoder, in order to learn high-level contextual semantics among sentences for style prediction.

To further improve the style coherence, the style embedding sequence $S_{-N},S_{1-N},\cdots,S_{-1}$ corresponding to the past sentences is also passed to the fusion context encoder.  
We merge the text-side and speech-side information to form a new token sequence, with a special unknown token ([UNK]) added to mark the style embedding to be predicted.
The token sequence $e$ is represented as:
\begin{gather}
    e=(C_{-N},C_{1-N},\cdots,C_{N},S_{-N},S_{1-N},\cdots,S_{-1},[UNK])
\end{gather}

\begin{figure}[!tb]
	\centering
	\includegraphics[width=0.70\linewidth]{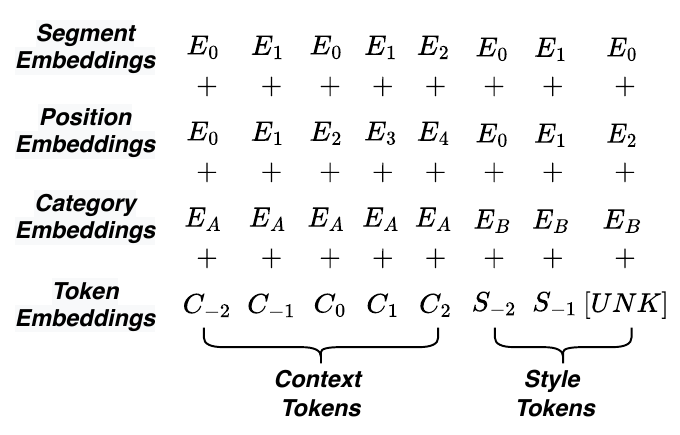}
	\caption{Context-aware Style Predictor input representation}
	\label{fig:input}
\end{figure}

To better learn the super-sentential prosody patterns, for each token, its input representation is constructed by summing the corresponding token, category, position, and segment embeddings.
A visualization of this construction when $N=2$ can be seen in Fig.\ref{fig:input}.
The category embeddings are utilized to distinguish text-side tokens from speech-side tokens.
The position embeddings are used to provide the relative position information of sentences.
And the segment embeddings are used to provide the sentence position in its paragraph.
Different from the sentence encoder, we specially design a mixture attention mask for the fusion context encoder inspired by \cite{dong2019unified}, to better utilize the context information and previous style information.
As shown in Fig.\ref{fig:predictor}, the context tokens only attend to each other context tokens from both directions, while the style tokens can attend to the previous style tokens and itself, as well as all the context tokens.
It is consistent with the process of human perception.
After encoding the two types of information, the output of context encoder is a sequence of context-sensitive representations, and the last representation (i.e., the representation corresponding to the special [UNK] token) is used as the predicted context-aware style embedding of the current sentence.

\subsection{Training Strategy and Inference Procedure}
During the training stage, to encourage the context-aware style predictor to learn style representations, our proposed model is trained with a knowledge distillation strategy in three steps.
In the first step, the acoustic model and the style extractor are jointly trained to get a well-trained style extractor in an unsupervised way.
In the second step, 
we use the style embedding extracted from
the ground-truth speech of the current sentence
as the target to guide the 
training of the predictor,
and the previous style embeddings are also extracted from the ground-truth.
Moreover, we apply the zero-padding method when the number of past or future sentences is less than $L$, both for text-side and speech-side.
Finally, we jointly train the acoustic model and the predictor with a lower learning rate to further improve the naturalness of synthesized speech.

During inference for multi-sentence in audiobook, the whole long-form speech is generated sentence by sentence and combined in order.
Since we don't have the ground-truth at this stage, the previous style embeddings can be extracted from previously synthesized speeches or other manually-determined speeches by the extractor.
In this way, our method can generate a long-form speech with contextually appropriate and coherent speaking style.

\section{Experiments}
\label{sec:exp}
\subsection{Experimental Setup}
For the experiments, we use an internal single-speaker audiobook corpus.
The corpus contains roughly 30 hours of recordings created by a professional male Mandarin native speaker reading about 87 chapters of a fiction with rich expressiveness.
The dataset has 14500 audio clips in total, of which 90\% of clips are used for training, and 10\% are used for validation and testing.

For feature extraction, we transform the raw waveforms into 80-dim mel-spectrograms with a sampling rate of 24 kHz.
A pre-trained Chinese XLNet-base model\footnote{\href{https://github.com/ymcui/Chinese-XLNet}{https://github.com/ymcui/Chinese-XLNet}} is used in our experiments.
Meanwhile, we conduct forced alignment using an automatic speech recognition tool to obtain phoneme duration.
The context considered in the context-aware style predictor is made up of the current sentence, its two past sentences, and its two future sentences.

We train models on a NVIDIA A10 GPU with a batch size of 16 up to 180k iterations for the first training step, 20k iterations for the second, and 30k iterations for the last, using Adam optimizer with $\beta_1=0.9$ , $\beta_2=0.98$.
The utterance encoder and context encoder both have 3 Transformer blocks.
In addition, a well-trained HIFI-GAN\cite{kong2020hifi} is used as the vocoder to generate the waveform.

In our evaluations, two FastSpeech 2 based baselines are implemented for comparison:

\textbf{FastSpeech 2:}
an open-source FastSpeech 2 \footnote{\href{https://github.com/ming024/FastSpeech2}{https://github.com/ming024/FastSpeech2}} \cite{fastspeech2}.

\textbf{HCE:}
hierarchical context encoder (HCE) \cite{lei2022towards} method, which predicts the style embedding from only context.

\subsection{Subjective Evaluation}
Since our proposed method is for audiobook speech synthesis, we conduct two kinds of mean opinion score (MOS) tests to evaluate the expressiveness and naturalness of the generated speech: 1) Single-sentence MOS (S-MOS): evaluate the synthesized speech of single sentence; 2) Multi-sentence MOS (M-MOS): evaluate the entire long-form speech which combines the synthesized speeches of multiple sentences in order.
A group of 20 native Chinese speakers are asked to listen to 15 samples and rate on a scale from 1 to 5 with 1-point interval.
As presented in Table \ref{tab:mos}, our proposed method achieves the best S-MOS of $3.990$ and M-MOS of $3.917$.
HCE gets the S-MOS result close to the proposed method, but there is a significant gap over $0.15$ with the proposed in M-MOS.
In addition, the proposed outperforms FastSpeech 2 greatly in both two metrics.

\begin{figure}[htbp]
	\centering
	\includegraphics[width=0.8\linewidth, height=0.35\linewidth]{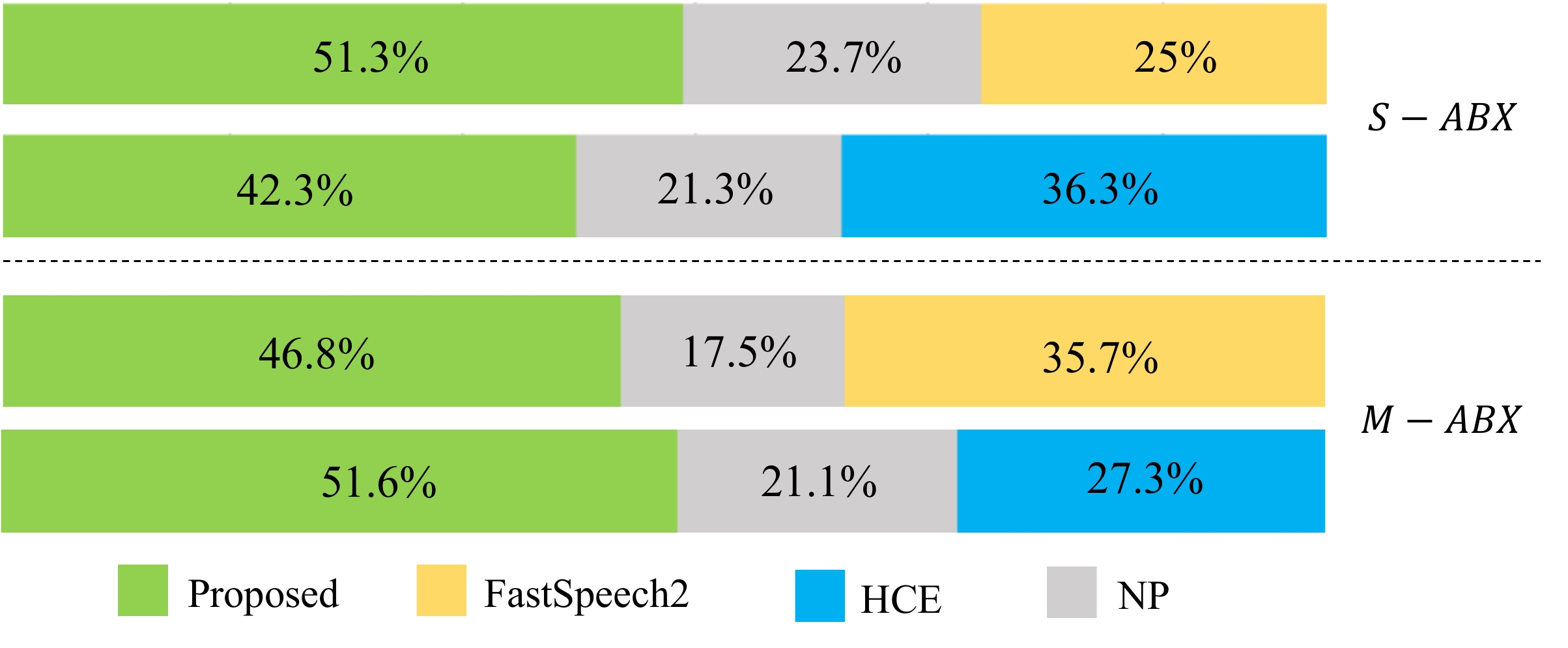}
	\caption{Results of the S-ABX and M-ABX preference tests. NP means no preference.}
	\label{fig:abx}
\end{figure}

ABX preference test is also conducted between our proposed method and the two baselines respectively.
Similarly, we conduct Single-sentence ABX (S-ABX) and multi-sentence ABX (M-ABX).
The preference test between two systems is to choose which one is preferable from the same perspective as the MOS test.
The results are shown in Fig.\ref{fig:abx}.
The preference rate of our proposed method exceeds FastSpeech 2 by $26.3\%$ for single-sentence and $11.1\%$ for multi-sentence, and exceeds HCE by $6.7\%$ for single sentence and $24.3\%$ for multi-sentence. 

\begin{table}[htbp]
\renewcommand{\arraystretch}{1.0}
  \caption{The S-MOS and M-MOS of different models with 95\% confidence intervals.}
  \label{tab:mos}
  \centering
  \begin{tabular}{l|c|c} 
    \toprule
    \textbf{Model} &\textbf{S-MOS} &\textbf{M-MOS}\\
    \midrule
    Ground Truth & $4.803\pm0.057$ & $4.826\pm0.060$~~~               \\
    FastSpeech 2 & $3.737\pm0.084$ & $3.792\pm0.081$~~~  \\
    HCE & $3.923\pm0.083$ & $3.761\pm0.078$~~~  \\
    Proposed & $\mathbf{3.990\pm0.072}$ & $\mathbf{3.917\pm0.078}$~~~ \\
    \bottomrule
  \end{tabular}
\end{table}
 
Both MOS and ABX preference tests demonstrate that our proposed method can improve the expressiveness and naturalness of synthesized speech not only for multi-sentence, but also for single-sentence.
For single-sentence, compared with the basic FastSpeech 2 that only uses phone sequence as input, our proposed method and HCE perform better, indicating that 
considering context information
can indeed help TTS model synthesize speech with more style variations.
The performance of our proposed model is further improved compared to HCE, because the super-sentential prosody patterns are taken into account.
In addition, we observe that speech synthesized by FastSpeech 2 has relatively smooth style transitions among sentences but lacks expressiveness, while HCE has rich expressiveness but varying greatly from one sentence to the next, both of which result in poor performance in the multi-sentence test.
This indicate that considering speech-side previous style information besides context is beneficial for generating expressive and coherent speech for audiobook speech synthesis.

\subsection{Objective Evaluation}
To measure the 
synthesis performance
objectively, we calculate the root mean square error (RMSE) of F0, the mean square error (MSE) of duration, and mel cepstral distortion (MCD) as the objective evaluation metrics followed by previous works \cite{fastspeech2, lei2022towards}.
Before calculating F0 RMSE, we first use dynamic time warping (DTW) to construct the alignment paths between the synthesized mel-spectrogram and the ground-truth.
Then, the F0 sequence extracted from the synthesized is aligned toward ground-truth following the DTW path.
We also utilize DTW to compute the minimum MCD by aligning the two mel-spectrograms.
For the duration, we calculate the MSE between the predicted and the ground-truth duration.
Moreover, to measure the accuracy of style prediction, we calculate the MSE of 
style embedding between the predicted and ground-truth.
The results are shown in Table \ref{tab:objective}.
Our proposed method outperforms the two baselines in all evaluation metrics, indicating that it can synthesize speech closer to the ground-truth.
Additionally, we also observe that the proposed considering speech coherence achieves further improvement in terms of style prediction compared to HCE. 

\begin{table}[!htb]\footnotesize
\renewcommand{\arraystretch}{1.0}
  \caption{Objective evaluations for different models.}
  \label{tab:objective}
  \centering
  \begin{tabular}{lccc} 
    \toprule
    \textbf{} & \textbf{FastSpeech 2} & \textbf{HCE} & \textbf{Proposed} \\
    \midrule
    F0 RMSE  & $65.1719$ & $63.6806$ & \textbf{62.1702} ~~~  \\
    Duration MSE & $0.2174$ & $0.2067$ & \textbf{0.1974} ~~~ \\
    MCD  & $5.0608$ & $5.0376$ & \textbf{4.8988} ~~~ \\
    Style MSE  & $-$ & $0.3148$ & \textbf{0.2524} ~~~ \\
    \bottomrule
  \end{tabular}
\end{table}

\subsection{Ablation Study}
We conduct three ablation studies to further investigate the effectiveness of several techniques used in our proposed model, including utilizing the mixture attention mask, previous style information, and hierarchical transformer architecture.
The comparison mean opinion score (CMOS) is employed to compare the naturalness and expressiveness of synthesized speech for multi-sentence text, and the style MSE is calculated to measure the accuracy of style prediction.
The results are shown in Table \ref{tab:cmos}.
Compared with the proposed method, 
these removals 
are degraded to various degrees respectively, which indicates that all these components substantially impact our proposed model.
Similar to HCE, the hierarchical architecture contributes greatly to style prediction.
Moreover, the result shows that introducing previous style information is helpful and our proposed mixture attention mask can further improve the performance.
\begin{table}[!htb]
\renewcommand{\arraystretch}{1.0}
 \caption{Results of ablation studies.}
 \label{tab:cmos}
 \centering
 \begin{tabular}{l|c|c} 
    \toprule
    \textbf{Model} & \textbf{CMOS} & \textbf{Style MSE}\\
    \midrule
    Proposed  & $-$ & \textbf{0.2524} ~~~  \\
    \hspace{3mm} - mixture attention mask  & -$0.148$ & $0.2670$~~~ \\
    \hspace{3mm} - previous style information  & -$0.172$ & $0.2805$ ~~~ \\
    \hspace{3mm} - hierarchical architecture  & -$0.220$ & $0.3255$ ~~~ \\
    \bottomrule
 \end{tabular}
\end{table}

\subsection{Case Study}
\begin{figure}[!htb]
	\centering
	\includegraphics[width=0.9\linewidth, height=0.6\linewidth]{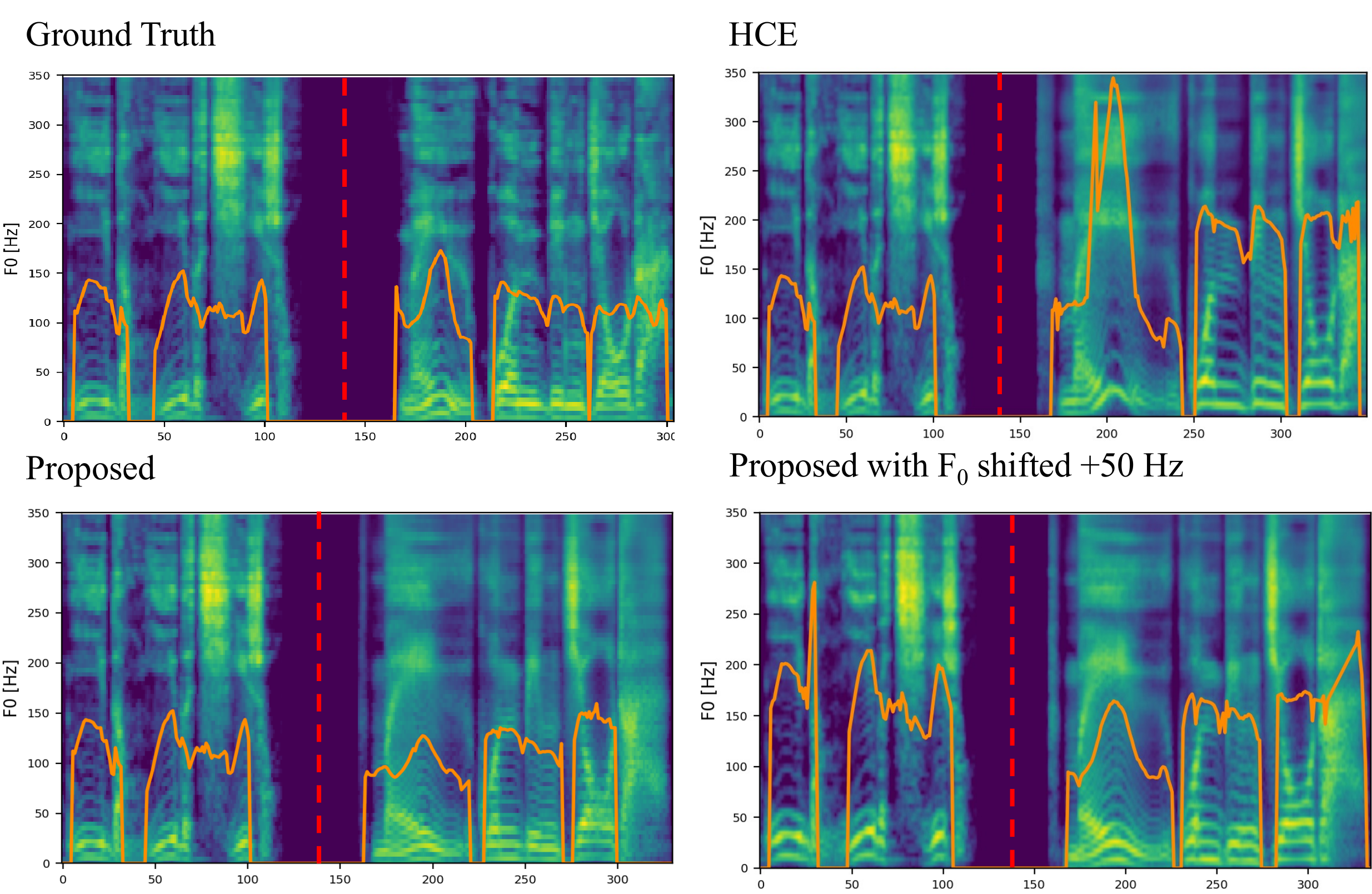}
	\caption{Mel-spectrograms and pitch contours of the speeches synthesized by different models for an example utterance in test set.}
	\label{fig:casestudy}
\end{figure}

As shown in Fig.\ref{fig:casestudy}, we make a case study by comparing the mel-spectrograms and pitch contours of the speeches generated from HCE and our proposed method, with the ground-truth previous speech concatenated.
Specifically, the red dashed line represents the sentence boundary, after which is the synthesized part.
It is observed that the speech synthesized by HCE contains larger pitch fluctuations than the proposed one.
But there exists a significant pitch variation from the previous utterance to the current due to the absence of previous style information, resulting in a certain degree of unnaturalness when combining these utterances to a long-form speech.
Compared with HCE, the speech synthesized by our proposed is more similar to the ground-truth speech in terms of intonation trends and style variations between the previous and the current utterance.
Moreover, 
we have tried to synthesize the same sentence by our proposed model but shifting the $F_0$ of the previous speech by $+50$ Hz.
It is observed that the $F_0$ of synthesized speech has also changed accordingly, indicating that our proposed model successfully learns to generate coherent style according to previous speech.
Thus, the same sentence can be synthesized in various ways to achieve more natural and expressive speech synthesis.

\section{Conclusions}
\label{sec:conclusions}
In this paper, 
we propose a context-aware coherent speaking style prediction method for audiobook speech synthesis.
A hierarchical transformer-based context-aware style predictor is introduced to consider both context information and style information of previous speeches, with a mixture attention mask.
Experimental results 
demonstrate that our proposed approach could significantly improve the expressiveness and coherence of the synthesized speech, for both single-sentence and multi-sentence test.

\vspace{+0.3cm}
\textbf{Acknowledgement}: This work is supported by National Natural Science Foundation of China (62076144), Major Key Project of PCL (PCL2021A06, PCL2022D01), Shenzhen Key Laboratory of next generation interactive media innovative technology (ZDSYS202106\\23092001004) and Shenzhen Science and Technology Program (WDZC20220816140515001, JCYJ20220818101014030).

\vfill\pagebreak
\ninept

\bibliographystyle{IEEEbib}
\bibliography{strings,refs}
\end{document}